%% file: PRL1.tex
\documentclass[aps,prl,twocolumn,showpacs,showkeys,superscriptaddress,groupedaddress]{revtex4}  % for review and submission
\usepackage{graphicx, epstopdf}  % needed for figures
\DeclareGraphicsExtensions{.ps}
\usepackage{dcolumn}   % needed for some tables
\usepackage{bm}        % for math
\usepackage{amssymb}   % for math
\usepackage{graphicx}

\begin{document}

% The following information is for internal review, please remove them for submission
%\widetext
%\leftline{Version xx as of \today}
%\leftline{Primary authors: Joe E. Physics}
%\leftline{To be submitted to (PRL, PRD-RC, PRD, PLB; choose one.)}
%\leftline{Comment to {\tt d0-run2eb-nnn@fnal.gov} by xxx, yyy}
%\centerline{\em D\O\ INTERNAL DOCUMENT -- NOT FOR PUBLIC DISTRIBUTION}

% the following line is for submission, including submission to the arXiv!!
%\hspace{5.2in} \mbox{Fermilab-Pub-04/xxx-E}

\title{A Processing In-Memory Realization Using QCA: Proposal and Implementation}
\input author_list.tex       % D0 authors (remove the first 3 lines
                             % of this file prior to submission, they
                             % contain a time stamp for the authorlist)
                             % (includes institutions and visitors)
%\date{\today}

\begin{abstract}
Processing in Memory (PIM) is a computing paradigm that promises enormous gain in processing speed by eradicating latencies in the typical von Neumann architecture. It has gained popularity owing to its throughput by embedding storage and computation of data in a single unit. We portray implementation of Akers array architecture endowed with PIM computation using Quantum-dot Cellular Automata (QCA). We present the proof of concept of PIM with its realization in the QCA designer paradigm. We illustrate implementation of Ex-OR gate with the help of QCA based Akers Array and put forth many interesting potential possibilities.
\end{abstract}



\maketitle{}

\section{I. Introduction}
%\section{\label{sec:level1}First-level heading}
% sections are not used for PRL papers

Disparity posed by the processing time by the processor versus data access speed is a bottleneck of the von-Neumann architecture which has apparent more in the recent backdrop of increasing requirement of computing power. Researchers are striving hard to address this bottleneck using various methods such as induction of cache memory, implementation of branch predictor algorithms, realization of morphware and configware [1]. In recent years, PIM architecture has gained popularity in the wake of increasingly complex application domains such as big data analytics, machine learning, soft computing and other emerging computing paradigms [2-3]. As against the requirement of two separate units for processing and storing of the data in the conventional von-Neumann, PIM architecture accomplishes the same only with one integrated unit. Therefore, PIM emerges as the winner in terms of speed, feature size and power consumption. Scholarly literature reveals various methods to implement the PIM architecture [4-7]. In this paper we present implementation of PIM using combination of Quantum Dot Cellular Automata (QCA) and Akers array. 

The Akers rectangular logic arrays, first proposed in 1972 by S. B. Akers [6] has a great developmental history. However, the same has been less traversed for the purpose of PIM computation. This kind of architecture has enormous potential with the implementation possibilities through nanotechnology which can be referred to as ‘nanoscale PIM’. The obvious advantages such as reduction in feature size leads to augmenting the computing speed as well as significant reduction the power consumption. In this regards, QCA is a promising and reliable technique as described by so many papers as the future electronics [8-11]. Since the Akers array is known for its PIM capabilities, we have integrated the same with QCA so as to synergize their wherewithal for improving the computing metrics. Realization of the same is depicted in this paper in a software environment of QCA designer suite. 

Rest of the paper is organized as follows, after general introduction; second section proposes the QCA Akers logic array. This is followed by evaluation of primitive logic cell and utilizing the same for forming an Ex-OR gate. Throughout the paper simulation results are presented. At the end conclusion and future work is reported. 

\section{II. Proposed QCA Akers logic array}
The proposed QCA-Akers logic array cells described herein are based on pioneering conception of Akers widely described in literature [10]. Every logic cell in the original design has three inputs with its output exhibiting the following function:
\begin{equation}
F(X, Y, Z) = X(\bar{Z}) +YZ	
\end{equation}

We propose to use the input variable ‘Z’ of Akers logic array to store the internal state of a modified Akers QCA cell. In this regards the inputs of the executed Boolean function are treated as the stored data in QCA cell. This facilitates the modified QCA Akers logic array to multiplex the functions viz. processing and storing the data. Using this functionality in the form of primitive, implementation of PIM is possible as depicted in the remainder of the paper. We further elaborate the structure and operations of logic cells in the following section

\section{III. Evaluation of primitive logic cell}
The basis of the proposed logic cell functionality is the equation (1). The primitive logic cell circuit structure is shown in fig. 1 (a). We have simulated the output of the above said logic structure using QCA Designer suite which is an freeware developed with the research effort by the Walus Group at the University of British Columbia for creating, designing and simulating designs based on Quantum Dot Cellular Automata (QCA) [12-13]. The digital simulation engine, one of the modules of the suite simulates the cells on the basis of null, logic 1 and logic 0 along with the appropriate clock stimulus, the system iterates till the convergence in terms of stable state is reached. The design is simulated as bistable simulation with the parameters set as default in the simulation suite as shown in table 1. The simulation outcome is shown in fig. 1 (b). The inputs of cell ‘X’ and ‘Y’ are given as fixed input i.e. zero and one respectively. The control input ‘Z’ is used for storing the logical state of Akers QCA cell. By using these control inputs the desired output from the Akers logic cell is derived. The designed circuit of primitive logic cell satisfies the output of Akers logic cell previously described in eq. 1.

\begin{figure}[h]
	\includegraphics[width=8cm]{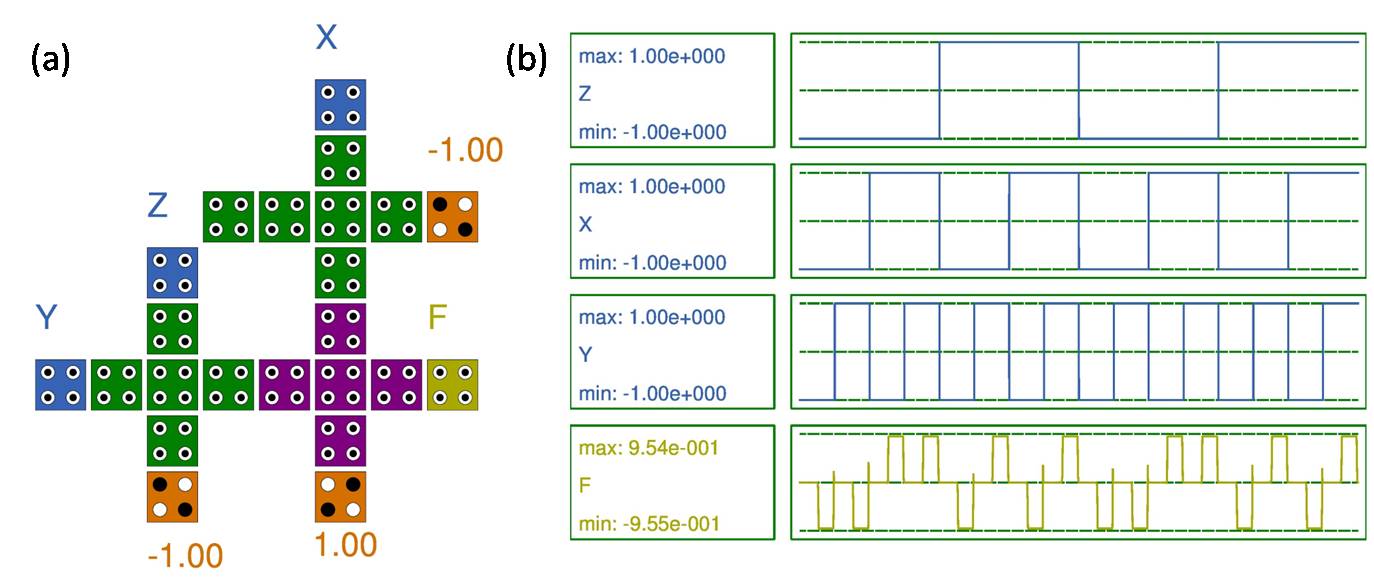}
	\caption{Logic cell. (a) Proposed primitive logic cell using QCA. (b) Its simulation results which gives output of eq. (1). }
\end{figure}

After establishing the functioning of the primitive logic cell and its confirmation through simulation we could implement basic logic gates by using the primitive. Since the logic gates can be used to form the complex logic systems, the intent is to evaluate their functionality.  We exemplify here implementation of Ex-OR gate owing to its hybrid nature.

\begin{table}[]
	\centering
	\caption{Simulation parameters set in the QCA Designer suite}
	\label{my-label}
	\begin{tabular}{|l|l|}
		\hline
		Parameter                       & Value           \\ \hline
		Temperature                     & 1 K             \\ \hline
		Relaxation Time                 & 1.000000e-015 s \\ \hline
		Time step                       & 1.000000e-015 s \\ \hline
		Clock High                      & 9.800000e-022 J \\ \hline
		Clock low                       & 3.800000e-023 J \\ \hline
		Clock Shift                     & 0.000000e+000   \\ \hline
		Clock Amplitude factor          & 2.000000        \\ \hline
		Radius of effect                & 80.000000 nm    \\ \hline
		Relative permittivity           & 12.900000       \\ \hline
		Layer separation                & 11.500000 nm    \\ \hline
		Convergence Tolerance           & 0.001000        \\ \hline
		Number of samples               & 128000          \\ \hline
		Maximum Integrations per sample & 100             \\ \hline
	\end{tabular}
\end{table}

%\textit{Physical Review} style requires that the initial citation of
%figures or tables be in numerical order in text, so don't cite
%Fig.~\ref{fig:wide} until Fig.~\ref{fig:epsart} has been cited.

\section{IV. Implementation Details of Processing In-memory Architecture }
In order to evaluate modified QCA Akers logic array, we consider Ex-OR gate as a representative candidate. The two input Ex-OR gate using QCA Akers logic array is shown in fig. 2. This Ex-OR gate structure is same as original Akers arrays Ex-OR gate but difference lies in the basic cell structure. One of the major drawbacks of the conventional Akers logic Ex-OR gate is that, the number of unit cells increases exponentially with the increase in the inputs [8]. As against the usage of primitive logic cell implemented herein results in considerably miniature size of QCA cells and therefore serves as the most apt choice for the next generation computing architecture.

\begin{figure}[h]
	\includegraphics[width=8cm]{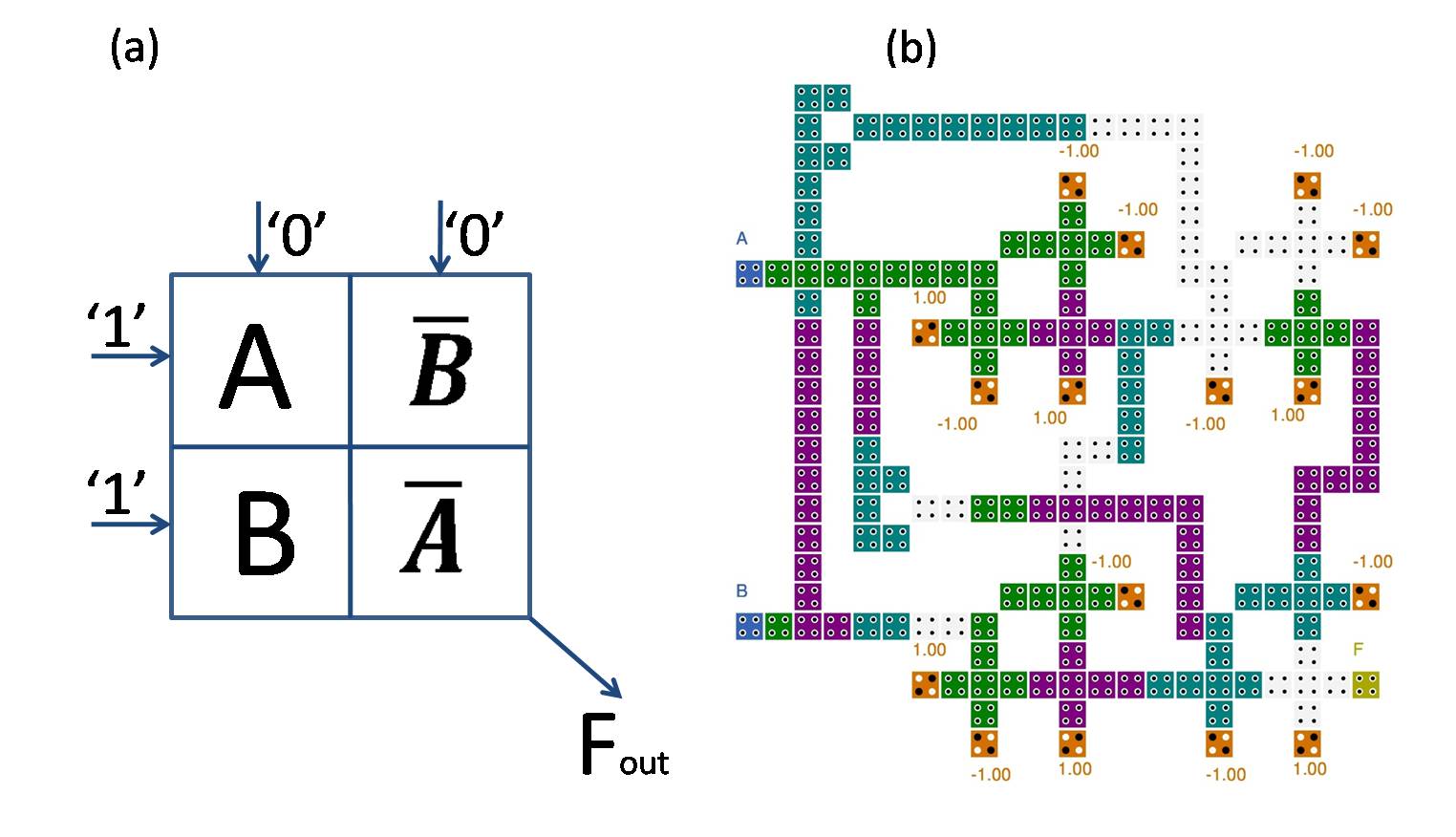}
	\caption{Two input Ex-OR gate. (a) Original Akers two input Ex-OR gate. (b)QCA Akers two input Ex-OR gate. }
\end{figure}

Two input QCA Akers logic Ex-OR gate consists of four Akers cells, each array consist of memory and computing part, as shown in fig. 2 (a). Two QCA cells circuits are connected in parallel manner as shown in fig. 2(b). The QCA Akers two input Ex-OR gate thus formed, was simulated in QCA Designer suite. The simulation parameters are given in table 1. Fixed inputs as per the Akers logic have been provided as shown in fig. 2(b), while various combinations were instantiated at ‘A’ and ‘B’. As per the inputs instantiated at ‘A’ and ‘B’, the diagonal position of electrons will remain constant till the next stimulus. This in fact implies the memory functionality. Moreover the memory functionality is non-volatile in nature since it does not get lost till the next instantiation of inputs through gated clock pulses. Using these QCA cells the logical output of each cell is controlled and it further acts as an input for neighbouring cell. 

Simulation output is shown in fig. 3 confirms the Ex-OR behaviour. Thus the QCA based Akers logic array exhibits Ex-OR logic functions. Moreover the control signal of QCA multiplexes the logic functionality along with the memory operation. Thus the implementation serves as one of the fundamental building blocks for the PIM architecture.

\begin{figure}[h]
	\includegraphics[width=8cm]{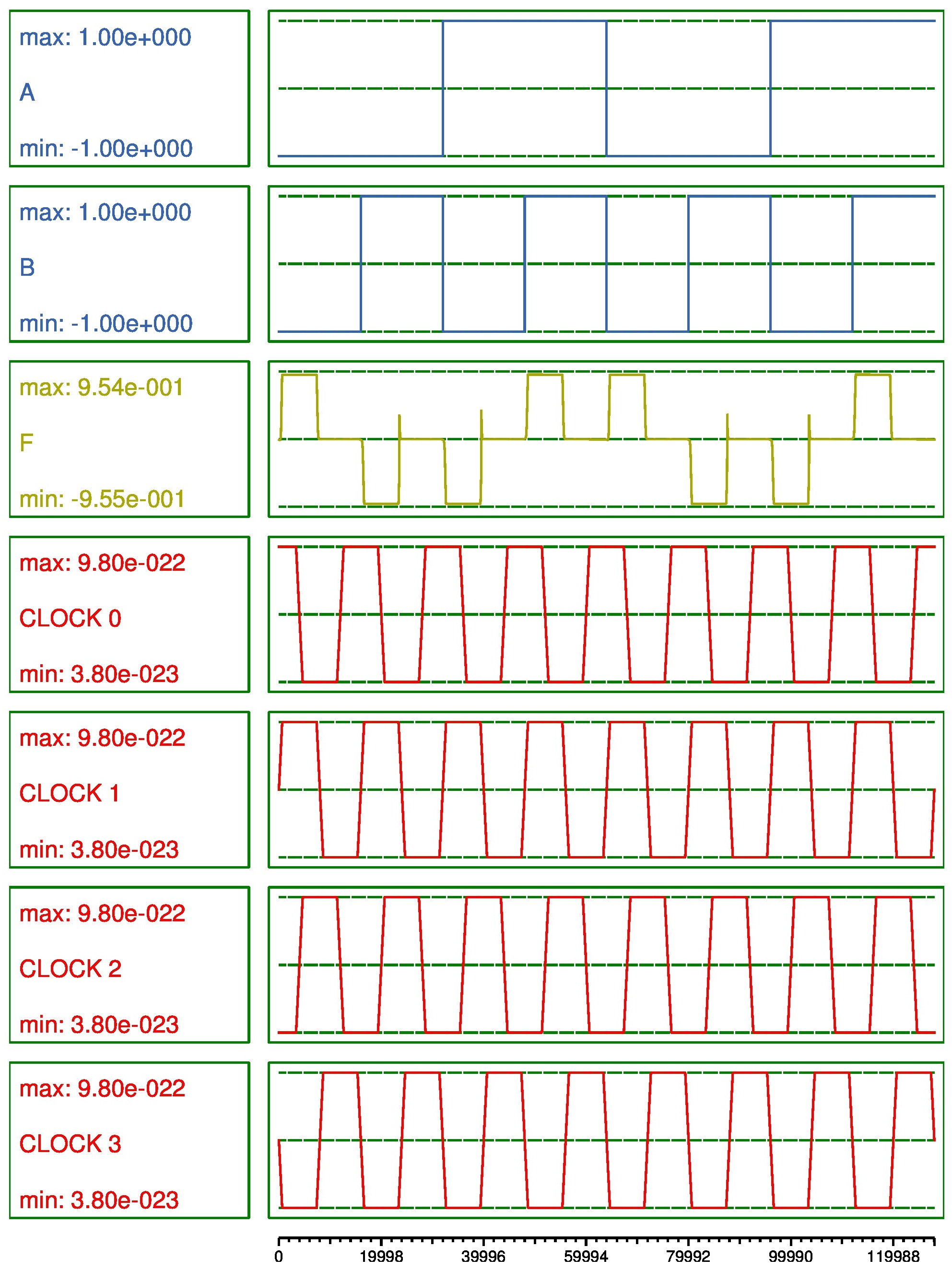}
	\caption{Simulation results of QCA Akers two input XOR gate.}
\end{figure}

\begin{table}[]
	\centering
	\caption{Performance and area parameters}
	\label{my-label}
	\begin{tabular}{|l|l|l|l|}
		\hline
		Design         & Cells & Area     & Clock Zone \\ \hline
		Primitive Cell & 23    & 0.04 $\mu m^2$ & 2          \\ \hline
		XOR gate       & 184   & 0.23 $\mu m^2$ & 19         \\ \hline
	\end{tabular}
\end{table}

\begin{table}[]
	\centering
	\caption{Power dissipation analysis of Ex-OR gate}
	\label{my-label}
	\begin{tabular}{|l|l|l|l|}
		\hline
		Parameter (meV)                                                                       & Ek =0.5 & Ek =1.0 & Ek =1.5 \\ \hline
		Max Kink Energy(meV)                                                                  & 0.00148 & 0.00148 & 0.00148 \\ \hline
		\begin{tabular}[c]{@{}l@{}}Max Energy dissipation\\  of circuit(meV)\end{tabular}     & 0.43081 & 0.49344 & 0.58046 \\ \hline
		\begin{tabular}[c]{@{}l@{}}Max Energy \\ dissipation vector\end{tabular}              & 03      & 03      & 03      \\ \hline
		\begin{tabular}[c]{@{}l@{}}Average Energy\\ dissipation of circuit(meV)\end{tabular}  & 0.23627 & 0.32811 & 0.44120 \\ \hline
		\begin{tabular}[c]{@{}l@{}}Max Energy dissipation\\ among all cells(meV)\end{tabular} & 0.00740 & 0.00729 & 0.00736 \\ \hline
		\begin{tabular}[c]{@{}l@{}}Max Energy \\ dissipation vector\end{tabular}              & 02      & 21      & 21      \\ \hline
		\begin{tabular}[c]{@{}l@{}}Min Energy dissipation\\ of circuit(meV)\end{tabular}      & 0.05694 & 0.17303 & 0.30943 \\ \hline
		\begin{tabular}[c]{@{}l@{}}Min Energy \\ dissipation vector\end{tabular}              & 11      & 11      & 11      \\ \hline
		\begin{tabular}[c]{@{}l@{}}Average Leakage\\ Energy dissipation(meV)\end{tabular}     & 0.05758 & 0.17475 & 0.31183 \\ \hline
		\begin{tabular}[c]{@{}l@{}}Average Switching\\ Energy Dissipation(meV)\end{tabular}   & 0.17869 & 0.15335 & 0.12937 \\ \hline
	\end{tabular}
\end{table}

\section{V. Conclusion and future work}
The QCA array presented in this paper comprises of QCA multiplexer circuit in every cell of Akers logic array. We proposed to multiplex this array as a memory and also for performing various Boolean operations. This unique attribute makes it a competitive candidate useful for the PIM computing applications. Table 2 shows the spatial complexity of the design. The Ex-OR gate realization was converged at much less complexity with 184 cells, 19 clock zones and with a footprint of 0.23 $\mu m^2$ which is significantly smaller than its CMOS counterpart.  Table 3 compares the power dissipation at different value of the kink energy. The kink energy is directly associated with the energy cost of the cells. The power dissipation is way smaller than the conventional CMOS counterpart in which it is of the order of few mWs. Thus the combination of QCA and Akers array provides many additional benefits over the conventional CMOS design. Moreover their synergic integration leads to the design with reduction in the power consumption and feature size, with improvement in the speed. We are in a process of extending the design to form a reconfigurable microprocessor.

\end{document}

%% file: author_list.tex
% remove these 3 lines before journal submittal.
\centerline{}
% end removal before journal submittal
%

%
\author{P.P.~Chougule} \affiliation{Computational Electronics and Nanoscience Research Laboratory,
School of Nanoscience and Biotechnology, Shivaji University, Kolhapur- 416004, India}
\author{B.~Sen} \affiliation{Department of Computer Science and Engineering,
National institute of Technology, Durgapur, W.B- 713209, India}
\author{R.~Mukherjee} \affiliation{Department of Computer Science and Engineering, 
National institute of Technology, Durgapur, W.B- 713209, India}
\author{V.C.~Karade} \affiliation{Computational Electronics and Nanoscience Research Laboratory,
School of Nanoscience and Biotechnology, Shivaji University, Kolhapur- 416004, India}
\author{P.S.~Patil} \affiliation{Computational Electronics and Nanoscience Research Laboratory,
School of Nanoscience and Biotechnology, Shivaji University, Kolhapur- 416004, India}
\author{T.D.~Dongale} \affiliation{Computational Electronics and Nanoscience Research Laboratory,
School of Nanoscience and Biotechnology, Shivaji University, Kolhapur- 416004, India}
\author{R.K.~Kamat} \affiliation{Embedded System and VLSI Research Laboratory, Department of Electronics, 
Shivaji University, Kolhapur, 416004, India}

%
% visitor_addresses.tex                       1 December 2015
%  available symbols are:
%  $\ast, \dag, \ddag, \S, \P, $\|$, $\ast\ast$, \dag\dag, \ddag\ddag ,\#
%